\documentstyle[12pt]{article}

\def\pconst{2^{\fr{1}{4}}}
\def\mconst{2^{-\fr{1}{4}}}
\def\pd{\partial}

\def\tpd{\partial_{\tau}}
\def\spd{\partial_{\sigma}}
\def\da{\dot{a}}
\def\db{\dot{b}}
\def\a{\alpha}
\def\ta{{\tilde \alpha}}

\def\s{\sigma}
\def\t{\tau}
\def\eps{\epsilon}

\def\lam{\lambda}

\def\Gm{\Gamma}
\def\gm{\gamma}

\def\momp{p^+}
\def\hdelta{{\hat \delta}}
\def\bdelta{{\bar \delta}}

\def\sq{\sqrt}
\def\twosq{\hbox{$\sqrt 2$}}
\def\e{{\rm e}}

\def\fr{\frac}
\def\tS{{\tilde S}}

\def\pp{\prime}

\def\bb{\begin{equation}}
\def\ee{\end{equation}}
\def\bba{\begin{eqnarray}}
\def\eea{\end{eqnarray}}

\begin{document}

\begin{titlepage}

\begin{tabbing}
   qqqqqqqqqqqqqqqqqqqqqqqqqqqqqqqqqqqqqqqqqqqqqq 
   \= qqqqqqqqqqqqq  \kill 
         \>  {\sc KEK-TH-522}    \\
         \>       hep-th/9706187 \\
         \>  {\sc June 1997} 
\end{tabbing}
\vspace{5mm}

\begin{center}
{\Large {\bf Supersymmetric Wilson Loops \break in IIB Matrix Model}}
\end{center}

\vspace{1.5cm}

\centering{\sc Ken-ji Hamada}\footnote{E-mail address : 
hamada@theory.kek.jp}

\vspace{1cm}

\begin{center}
{\it National Laboratory for High Energy Physics (KEK),} \\
{\it Tsukuba, Ibaraki 305, Japan}
\end{center} 

\vspace{1cm}

\begin{abstract} 
We show that the supersymmetric Wilson loops in IIB matrix model give a 
transition operator from reduced supersymmetric Yang-Mills theory to 
supersymmetric space-time theory. 
In comparison with Green-Schwarz superstring we identify the supersymmetric 
Wilson loops with the asymptotic states of IIB superstring. 
It is pointed out that the supersymmetry transformation law of the 
Wilson loops is the inverse of that for the vertex operators of  
massless modes in the $U(N)$ open superstring with Dirichlet boundary 
condition. 
\end{abstract}

\end{titlepage}

\section{Introduction}
\indent

   For a long time it has been hoped that the large $N$ gauge theory~\cite{t} 
will give a nonperturbative definition of string theory. 
In the begining of '90 2D string has solved exactly in 
terms of matrix models~\cite{bk} and many works have been carried out 
in this field~\cite{bw}.  
The identification with the continuum theory was done in the 
direct calculations of amplitudes~\cite{gl} and then was completed  
using the $W_{\infty}$ symmetry~\cite{ha}.  

   Recently more realistic matrix models called M(atrix) theory~\cite{bfss} 
and IIB matrix theory~\cite{ikkt} (and also see [8 -- 15]) have been proposed, 
which are described in terms of the D-particles and 
the D-instantons~\cite{p,gg}. 
In this paper we study the IIB matrix model, which is hoped to give 
the type IIB superstring.   
In this case, other than 2D string, the oscillation modes 
with continuous momenta will arise. 
The aim of this paper is to clarify how the oscillation modes arise 
in the IIB matrix theory. We study such an issue using the supersymmetry. 

  The Wilson loops will describe the operators which create and annihilate 
strings~\cite{ikkt,fkkt}. We here introduce supersymmetric Wilson loops in 
IIB matrix theory and identify it with the asymptotic state of superstring. 
To carry out the program we first study the supersymmetry transformation 
law of the wave function of IIB superstring which is constructed by acting the 
vertex operator of the D-instanton~\cite{hb} on the boundary state~\cite{gg}.    
We then show that the supersymmetric Wilson loop just has the same 
property as that the state of superstring has, where the supersymmetry 
transformation of reduced super Yang-Mills theory acts on it as a counterpart 
of that of world-sheet theory.

\section{The state of IIB superstring}
\setcounter{equation}{0}
\indent

   We first construct the eigenstate of Hamiltonian using Green-Schwarz 
superstring quantized in the light-cone gauge~\cite{gsb,gsw} 
and then discuss its supersymmetry transformation law. 

 Let us consider cylinder frame with Dirichlet boundary at $\t=0$. 
The boundary state is defined by the conditions 
\bb
    \spd X^{\mu}(\s)|B> = 0 ~, \qquad  S^{+a}(\s)|B> = 0 ~,
\ee
where $S^{\pm a}= \fr{1}{\sq 2}( S^a \pm i\tS^a )$ and $\mu = (+,-, I) \quad 
I=1, \cdots ,8$. This is the D-instanton state 
discussed in~\cite{gg,hb}. In the following we mainly use the notations and 
conventions of ref.~\cite{hb}. 
The conditions can be solved easily and we obtain 
\bb
    |B> = \exp \biggl[ \sum^{\infty}_{n=1} 
                 \biggl( \fr{1}{n}\a^I_{-n} \ta^I_{-n} 
                         -i S^a_{-n} \tS^a_{-n} \biggr) 
               \biggr] |B_0> ~,
\ee
where $|B_0>=|I>|I> -i|\da>| \da>$. The mode expansions of string 
coordinates are defined by
\bba 
  &&  X^I (\t,\s) = x^I + p^I \t 
            + \fr{i}{2}\sum_{n \neq 0} \fr{1}{n} 
            \Bigl( \a^I_n \e^{-2in\pi(\t-\s)}
                     +\ta^I_n \e^{-2in\pi(\t+\s)} \Bigr) ~, 
           \nonumber  \\
  && S^a(\t,\s) = \sum_n S^a_n \e^{-2in\pi(\t-\s)} ~,  
                      \\
  && \tS^a(\t,\s) = \sum_n \tS^a_n \e^{-2in\pi(\t+\s)} ~. 
             \nonumber 
\eea     

  The vertex operator of (single) D-instanton~\cite{hb} is defined in terms of 
the broken currents $\tpd X^{\mu}$ and $S^-$ for translational invariance 
and supersymmtery in the form 
\bb
   V \bigl(x^{\mu}(\s),\eta(\s) \bigr) 
    = \int^{\pi}_0 \fr{d\s}{\pi} \Bigl\{ 
           x_{\mu}(\s) - i {\bar \eta}(\s)\Gm_{\mu}\theta
                     \Bigr\} \tpd X^{\mu} ~,  
\ee
where $ \theta = (\pconst 2i \sq{\momp})^{-1}\Gm^+ S^-$. We here 
consider the $\s$-dependent functions $x^{\mu}(\s)$ and $\eta(\s)$. 
In the light-cone gauge defined by  
$x^+(\s)= x^+ = \t$ and $\Gm^+ \eta=0$, the vertex operator reduces to 
the simple form 
\bb
   V \bigl(x^{\mu},\eta \bigr) 
    = \int^{\pi}_0 \fr{d\s}{\pi} \Bigl\{ 
           x^I (\s)\tpd X^I - x^-(\s) \momp 
               + i \mconst \hbox{$\sq{\momp}$}\eta^a(\s)S^{-a} 
                     \Bigr\} ~.   
\ee
  
  Let us consider the Wilson loop operator $w =\exp(-iV)$ and act 
it on the boundary state. Using the Baker-Campbell-Hausdorff formula we can  
obtain the following state: 
\bba
   &&  |x,\eta> = w |B>   
                 \nonumber \\ 
   && ~ = \exp \Bigl(-ix^I_0 p^I +ix^-_0 \momp 
                  +\mconst \hbox{$\sq{\momp}$}\eta^a_0 S^{-a}_0 \Bigr)
                 \nonumber \\ 
   && \quad \times 
                 \exp \biggl[ \sum^{\infty}_{n=1}\biggl\{ 
                                 - n x^I_n x^I_{-n} 
                              -2i \Bigl( x^I_n \a^I_{-n} 
                                       +x^I_{-n}\ta^I_{-n} \Bigr)
                              +\fr{1}{n}\a^I_{-n}\ta^I_{-n} \Biggr\} 
                                 \biggr]  
                        \\
   && \quad \times 
                  \exp \biggl[ \sum^{\infty}_{n=1}\biggl\{ 
                     \fr{\momp}{2\twosq} \eta^a_n \eta^a_{-n} 
                      + \pconst \hbox{$\sq{\momp}$} 
                   \Bigl( \eta^a_n S^a_{-n} -i \eta^a_{-n}\tS^a_{-n} \Bigr)
                     -i S^a_{-n} \tS^a_{-n} \biggr\} \biggr] |B_0> ~, 
                    \nonumber
\eea
where the mode expansions of $x$ and $\eta$ are defined by
\bb
    x^I(\s)=\sum_n x^I_n \e^{2in\s} ~, \qquad 
    \eta^a(\s)=\sum_n \eta^a_n \e^{2in\s} ~.
\ee
This state satisfies the boundary conditions $X^I(\s)|x,\eta>=x^I(\s)|x,\eta>$ 
and $S^{+a}(\s)|x,\eta>=-2^{-\fr{1}{4}} \hbox{$\sq{\momp}$}\eta^a(\s)|x,\eta>$. 
{}From the expression of vertex operator, the operators $\tpd X^I(\s)$ and 
$S^{-a}(\s)$ are described in terms of the functional derivatives of $x^I(\s)$ 
and $\eta^a (\s)$, respectively.  

  The light-cone Hamiltonian is given by
\bb
   H = \fr{1}{2\momp} \int^{\pi}_0 \fr{d\s}{\pi} 
          \Bigl[ \bigl( \tpd X^I \bigr)^2 +\bigl( \spd X^I \bigr)^2  
                  -iS^{+a} \spd S^{+a} -i S^{-a}\spd S^{-a} \Bigr] ~.
\ee
Therefore the state $|x,\eta,\t> =\e^{i\t H}|x,\eta>$ satisfies the 
Schr\"{o}dinger equation
\bb
   -i \fr{\pd}{\pd \t}|x,\eta,\t> = H |x,\eta,\t> = h |x,\eta,\t> 
\ee 
where 
\bba
   && h = \fr{1}{2\momp} \int^{\pi}_0 \fr{d\s}{\pi} 
            \biggl[ -\pi^2 \biggl( \fr{\delta}{\delta x^I(\s)} \biggr)^2 
                   +\bigl( \spd x^I \bigr)^2  
             \nonumber     \\
   && \qquad\qquad
              -i \fr{\twosq \pi^2}{\momp} \fr{\delta}{\delta \eta^a(\s)} 
                            \spd \fr{\delta}{\delta \eta^a(\s)} 
                -i\fr{\momp}{\twosq} \eta^a(\s) \spd \eta^a(\s) 
                            \biggr]  
\eea
and the wave function is defined by $\Phi^{\star}(x,\eta,\t)=<\Phi |x,\eta,\t>$. 
 
  In the last of this section we discuss supersymmetry transformation law 
of the state. 
The supersymmetry transformation of vertex operator with respect to the 
unbroken supercharge $Q^+$ is translated into the supersymmetry 
transformation on $x^{\mu}$ and $\eta$. 
Using the equations
\bba
  && \hdelta^{(+)}_{\a}(\tpd X^I)=\pconst \fr{i}{\sq{\momp}} 
             \a^{\da}\gm^I_{a\da}\spd S^{-a} ~, 
             \nonumber  \\ 
  && \hdelta^{(+)}_{\a} S^{-a}= \pconst \hbox{$\sq{2\momp}$}\a^a 
          +\pconst \fr{1}{\sq{\momp}}\a^{\da}\gm^I_{a\da}\tpd X^I ~, 
\eea 
where $\hdelta^{(+)}_{\a}=[\a^a Q^{+a} +\a^{\da} Q^{+\da}, \quad]$, 
we obtain the equation  
\bb
    \hdelta^{(+)}_{\a} V(x^{\mu},~\eta) 
          = V(\bdelta_{\a}x^{\mu},~\bdelta_{\a}\eta ) ~, 
\ee
where 
\bba
   && \bdelta_{\a}x^I(\s) =-i \a^{\da} \gm^I_{a\da}\eta^a(\s) ~, 
               \nonumber \\
   &&  \bdelta_{\a}x^-_0 =i\twosq \a^a \eta^a_0 ~, 
               \label{eq:transf-c} \\ 
   &&  \bdelta_{\a}\eta^a (\s)
          = -\fr{\twosq}{\momp}\a^{\da}\gm^I_{a\da}\spd x^I(\s) ~. 
              \nonumber
\eea 
Using this we obtain $\hdelta^{(+)} w =\bdelta w$. 
Thus the Wilson loop gives the transition operator from  
the world-sheet theory to the space-time theory.

\section{Supersymmetric Wilson loops in IIB matrix model}
\setcounter{equation}{0}
\indent
 
   The world-sheet is regulated in the large $N$ picture. 
The reduced supersymmetric Yang-Mills theory will play an fundamental role 
to make a world-sheet. The identification of gauge fields with  space-time 
coordinates will gives a non-pertubative definition of the type IIB 
superstring~\cite{ikkt}. 
The states in IIB matrix model should have the same supersymmetry 
transformation law as that the continuum theory has.  
In this section we show that the supersymmetric Wilson loop just has 
the expected property. So we identify it with the IIB superstring state 
(in momentum space). 
 
  The supersymmetric Wilson loop operator we introduce here is    
\bb
     w(C) = tr \prod^M_{j=1} U_j ~, \qquad U_j = \e^{-i\eps V_j} ~, 
\ee
where $V_j$ is defined using the superfield in the form  
$V_{j} = k^{\mu}_j {\cal A}_{\mu}(\lam_j)$, where ${\cal A}_{\mu}(\lam_j)
 = \e^{{\bar \lam}_j G} A_{\mu}\e^{-{\bar \lam}_j G}$. 
$G$ is the generator of the supersymmetric Yang-Mills transformation 
\bb
     \delta_{\a} A_{\mu} = i{\bar \a}\Gm_{\mu}\Psi ~, \qquad 
     \delta_{\a} \Psi = -\fr{i}{2}[A_{\mu} , A_{\nu}]\Gm^{\mu\nu}\a ~.
\ee
Thus we define 
\bba
    &&  V_j = k^{\mu}_j \Bigl( A_{\mu} -i{\bar \Psi} \Gm_{\mu} \lam_j 
             + \fr{1}{4} [A_{\nu} ,A_{\lam}] 
                    {\bar \lam}_j \Gm_{\mu}\Gm^{\nu\lam} \lam_j 
                 \nonumber \\ 
    && \qquad\qquad\qquad 
              - \fr{i}{6}[ A_{\nu} ,{\bar \Psi}]\Gm_{\lam}\lam_j 
                     {\bar \lam}_j \Gm_{\mu}\Gm^{\nu\lam}\lam_j 
              + \cdots \Bigr) ~.
\eea

 In the following we work in the light-cone gauge~\footnote{
The covariant description of supersymmetry does not go well, where 
the constraint equation which serves as eq. (\ref{eq:constraint}) is not known.}
%%%%%%%%%%%%%%%%%%%%%%%%%   
\bb 
     k^+_j =k^+ ~, \qquad \Gm^+ \lam_j =0 ~.  
\ee
Then we can see that the following supersymmetry transformation is 
realized: 
\bb
        \delta_{\a} w(C) = \bdelta_{\a} w(C) ~,
\ee
where
\bba
    && \bdelta_{\a}k^I_j = 2i \a^{\da} \gm^I_{a\da}\Delta \lam^a_j ~,
           \nonumber \\
    && \bdelta_{\a} k^+ =0 ~, 
           \label{eq:transf-mom}            \\
    && \bdelta_{\a} \lam^a_j = \a^a + \fr{1}{\twosq k^+} 
                    k^I_j \gm^I_{a\da} \a^{\da} 
                \nonumber
\eea
and $\Delta \lam^a_j = \fr{1}{\eps}(\lam^a_{j+1}-\lam^a_j )$. 
The transformation of $k^-_j$ is defined through the equation  
\bb
     k^-_j = \fr{1}{2k^+}\bigl( k^I_j \bigr)^2  
                - i\twosq \lam^a_j \Delta \lam^a_j  
             \label{eq:constraint}
\ee
in the form $\bdelta_{\a} k^-_j =-i\twosq \a^a \Delta \lam^a_j 
-i \Delta (\fr{1}{k^+} k^I_j \lam^a_j \gm^I_{a\da} \a^{\da} )$. 
In the continuum limit $k^{\mu}_j \rightarrow k^{\mu}(\s)$ and 
$\Delta \lam^a_j \rightarrow \spd \lam^a (\s)$, 
this just corresponds to the supersymmetry transformation in momentum space 
derived in the continuum theory (\ref{eq:transf-c})~\footnote{ 
The transition function between coordinate space to momentum space is 
given by $w_f = \exp \{ i (-x^+ k^-_0 -x^-_0 k^+ 
+\int \fr{d\s}{\pi} x^I (\s)k^I(\s) +i\twosq k^+ \int \fr{d\s}{\pi} 
\eta^a(\s)\lam^a(\s) ) \}$ such that 
$\bdelta^{(c)} w_f=\bdelta^{(m)} w_f$, where $\bdelta^{(c)}$ and 
$\bdelta^{(m)}$ are defined by (\ref{eq:transf-c}) and 
(\ref{eq:transf-mom}), respectively.}. 
%%%%%%%%%%%%%%%%%%%%%%%%%%%%%%%%
In this case the supersymmetric Yang-Mills transformation $\delta$ 
just corresponds to the variation of world-sheet theory $\hdelta^{(+)}$. 
The constraint (\ref{eq:constraint}) corresponds to the boundary 
condition $\tpd X^- |B> = \fr{1}{2\momp}[(\tpd X^I)^2  
-i S^{-a}\spd S^{-a}] |B> $.

   The above transformation law can be proved in the following.  
In the light-cone gauge the matrix $V_j$ is described in $SO(8)$ notation 
as    
\bb
     V_j = V_j^0 +V_j^1 + V_j^2 + V_j^3 + \cdots ~, 
\ee
where
\bba
  && V_j^0 = k^I_j A^I -k^+ A^- -k^-_j A^+  ~,
              \nonumber  \\
  && V_j^1 = -i\twosq k^+ \lam^a_j \psi^a  
             -i\twosq k^I_j \lam^a_j \gm^I_{a\da} \psi^{\da}  ~, 
              \nonumber  \\ 
  && V_j^2 = \fr{k^+}{2\twosq}[A^I, A^J]
                    \lam^a_j \gm^{IJ}_{ab}\lam^b_j 
              - \fr{1}{2\twosq}k^I_j [A^+ ,A^J] 
                      \lam^a_j (\gm^I \gm^J)_{ab} \lam^b_j ~, 
                     \\  
  && V_j^3 = - \fr{i}{3\twosq}k^+ [A^I , \psi^{\da}\gm^J_{a\da} \lam^a_j ] 
                     \lam^b_j \gm^{IJ}_{bc} \lam^c_j 
             + \fr{i}{3\twosq}k^I_j [A^+ , \psi^{\da}\gm^J_{a\da} \lam^a_j ] 
                     \lam^b_j (\gm^I \gm^J)_{bc} \lam^c_j  ~. 
               \nonumber
\eea
The supersymmetry transformation is described in the 
$SO(8)$ notation as 
\bba
   && \delta_{\a} A^I = -i\a^{\da} \gm^I_{a\da}\psi^a 
                         -i\a^a \gm^I_{a\da}\psi^{\da} ~,
              \nonumber \\
   && \delta_{\a} A^+ = i \twosq \a^{\da} \psi^{\da} ~,
              \nonumber \\
   && \delta_{\a} A^- = i \twosq \a^a \psi^a ~,
                        \\
   && \delta_{\a} \psi^a = \fr{i}{2}[A^I , A^J] \gm^{IJ}_{ab}\a^b 
                           - i\twosq [A^- , A^I] \gm^I_{a\da}\a^{\da} 
                           + i[A^+ , A^-] \a^a ~, 
              \nonumber \\
   && \delta_{\a} \psi^{\da} = \fr{i}{2}[A^I , A^J] \gm^{IJ}_{\da\db}\a^{\db} 
                           - i\twosq [A^+ , A^I] \gm^I_{a\da}\a^a 
                           - i[A^+ , A^-] \a^{\da} ~.
              \nonumber 
\eea

  Let us first consider the variation of $V^0_j$ under the supersymmetry 
transformation $\delta$. 
We can easily obtain the following equation:
\bb
    \delta_{\a} V^0_j = \bdelta_{\a}(V^0_j + V^1_j ) + \Delta f^0_j ~, 
\ee
where
\bb
     f^0_j = -2i \a^{\da}\gm^I_{a\da} \lam^a_j A^I 
              - i\twosq \a^a \lam^a_j A^+  
              + \fr{i}{k^+}k^I_j \a^{\da} \gm^I_{a\da}\lam^a_j A^+ ~.  
\ee
In the next step we obtain 
\bb
    \delta_{\a} (V^0_j + V^1_j)  
       = \bdelta_{\a}(V^0_j + V^1_j + V^2_j ) + Y^0_j 
           + \Delta (f^0_j +f^1_j) ~, 
\ee
where
\bb
     Y^0_j = i[f^0_j ,V^0_j]    \label{eq:y-term}
\ee
and
\bb
      f^1_j = i\fr{\twosq}{3}[A^+ ,A^J] \a^{\da}\gm^I_{a\da}\lam^a_j 
                    \lam^b_j (\gm^I \gm^J)_{bc}\lam^c_j ~.
\ee
In general we will obtain the equation
\bb
        \delta_{\a} V_j = \bdelta_{\a}V_j + Y_j + \Delta f_j ~, 
              \label{eq:general-eq}
\ee
where $Y_j =Y^0_j + Y^1_j + \cdots$ and $f_j = f^0_j + f^1_j + \cdots$. 

   We will see that the extra term $Y_j$ is canceled in the Wilson loop.  
Let us consider the supersymmetry transformation of the Wilson loop.  
Using the relation (\ref{eq:general-eq}) we obtain 
\bba
   \delta_{\a} w(C) 
    & =& - tr \sum^M_{l=1} \biggl( \prod^{l-1}_{j=1}U_j \biggr)
              i\eps \delta_{\a} V_l 
                 \biggl( \prod^M_{j=l}U_j \biggr) 
            \nonumber \\
    & =& - tr \sum^M_{l=1} \biggl( \prod^{l-1}_{j=1}U_j \biggr)
            i\eps \Bigl( \bdelta_{\a}V_l +Y_l +\Delta f_l \Bigr)
                 \biggl( \prod^M_{j=l}U_j \biggr) ~. 
\eea
Noting that $\Delta f_l = \fr{1}{\eps}(f_{l+1}-f_l)$ and $f_l$ is the 
matrix such that   
$\fr{1}{\eps}f_l U_{l-1} = U_{l-1} (\fr{1}{\eps}f_l - i[f_l , V_{l-1}] 
+o(\eps) )$, we get the following expression: 
\bb
   \delta_{\a} w(C) 
     = - tr \sum^M_{l=1} \biggl( \prod^{l-1}_{j=1}U_j \biggr)
            i\eps \Bigl( \bdelta_{\a}V_l +Y_l - i[f_l ,V_{l-1}] \Bigr)
                 \biggl( \prod^M_{j=l}U_j \biggr) ~. 
\ee
As shown in the above calculation (\ref{eq:y-term}), $Y_l$  cancels 
$i[f_l , V_{l-1}]$ iteratively in the continuum limit. This cancellation 
is an analogy of that by contact terms in open superstring with 
Chan-Paton factor~\cite{hb,gs}. 
Thus we can prove the supersymmetry transformation law of the Wilson loop. 

  This is the inverse picture of supersymmetry transformation law of 
the vertex operator for massless mode in the $U(N)$ Dirichlet open superstring 
derived in~\cite{hb}, where the roles of the 
world-sheet theory and space-time theory are exchanged. 
Supersymmetric Yang-Mills theory now plays an role of world-sheet theory, 
not of the space-time one.

\section{The S-matrix}
\setcounter{equation}{0}
\indent

   In the previous section we discussed the supersymmetric Wilson loop in 
IIB matrix theory. We proposed that, in the symmetrical point of view, 
it corresponds to the asymptotic state of IIB superstring. The correlation 
function of the Wilson loops is defined by 
\bb
     <w(k_1,\lam_1) \cdots w(k_L,\lam_L)> 
       = \int dA d\Psi w(C_1) \cdots w(C_L) \exp (-S) 
\ee
where $S$ is the reduced supersymmetric Yang-Mills acton. The continuum limit is 
defined by $M\eps =1$ and $g^2 N =1$, where $g$ is the gauge coupling behaived 
as $g \sim \eps$.   
The momentum conservation comes from the integration over the $U(1)$ part in 
$U(N)$ matrix. The $U(1)$ part of $A^-$ integral gives the delta function 
$\delta (k^+_1 + \cdots + k^+_L)$ and others gives 
$\delta ( \eps\sum^M_{j=1}k^{\mu}_{1j} + \cdots 
+ \eps\sum^M_{j=1}k^{\mu}_{Lj})$, where $\mu=-,I$. 
In the continuum limit these give the momentum conservations of zero-modes,  
$k_0^{\mu}=\int \fr{d\s}{\pi} k^{\mu}(\s)$.  

Thus the $S$-matrix is defined by attaching the wave functions of 
oscillation modes $\Phi(k,\lam)$ in the form:
\bb
    S_{i \rightarrow f} 
        = \int \prod^L_{q=1}\fr{1}{\sq{|k^+_q|}} 
             [D^{\pp}k^I_q][D\lam^a_q] \Phi(k_q,\lam_q) 
                   < w(k_1,\lam_1) \cdots w(k_L,\lam_L)> 
\ee
where $[D^{\pp}k^I ][D \lam^a]$ means the integration over transverse 
oscillation modes and the prime stands for the exclusion of the zero-modes. 
Incoming states (outgoing states) are defined by the Wilson loops 
with $k^+$ positive (negative).   

  Finally we briefly comment on the Schwinger-Dyson equation of the 
followin type: 
\bb
     \int dA d\Psi \fr{\pd}{\pd A^{+ \a}} 
          \sum^M_{l=1}tr \biggl( \prod^{l-1}_{j=1} U_j ~ t^{\a}~  
                 \prod^M_{j=l} U_j \biggr) \exp(-S)  =0 ~. 
\ee
This is likely to correspond to the equation $<0| H |\Phi>=0$ in the continuum 
theory. Here the effects of the terms corresponding to $(\spd x^I)^2$ and 
$\eta^a \spd \eta^a$ in the Hamiltonian will be included in the derivative 
of action with respect to $A^+$. This is similar to the picture of the 
Hartle-Hawking wave function.

\begin{flushleft}
{\bf Acknowledgements}
\end{flushleft}

    I wish to thank our colleagues at KEK,  
especially N. Ishibashi, H. Kawai and A. Tsuchiya, for discussions.

\end{document}